\documentstyle [prl,aps,floats,twocolumn]{revtex}
\input psfig.tex
\begin{document}

\def\ba{\begin{eqnarray}}
\def\ea{\end{eqnarray}}
\def\be{\begin{equation}}
\def\ee{\end{equation}}
\def\({\left(}
\def\){\right)}
\def\[{\left[}
\def\]{\right]}
\def\lagrange {{\cal L}}
\def\del {\nabla}
\def\d {\partial}
\def\Tr{{\rm Tr}}
\def\half{{1\over 2}}
\def\fourth{{1\over 8}}
\def\bibi{\bibitem}
\def\S{{\cal S}}
\def\xx{\mbox{\boldmath $x$}}
\newcommand{\labeq}[1] {\label{eq:#1}}
\newcommand{\eqn}[1] {(\ref{eq:#1})}
\newcommand{\labfig}[1] {\label{fig:#1}}
\newcommand{\fig}[1] {\ref{fig:#1}}
\newcommand\bigdot[1] {\stackrel{\mbox{{\huge .}}}{#1}}
\newcommand\bigddot[1] {\stackrel{\mbox{{\huge ..}}}{#1}}

\twocolumn[\hsize\textwidth\columnwidth\hsize\csname @twocolumnfalse\endcsname
\title{Polarization of
the Microwave Background in Defect Models} \author{
Uro\v s Seljak \thanks{email:useljak\@cfa.harvard.edu},
Ue-Li Pen \thanks{email:upen\@cfa.harvard.edu}, and Neil
Turok\thanks{email:N.G.Turok\@amtp.cam.ac.uk}}
\address{${}^*$ Harvard-Smithsonian Center for Astrophysics,
60 Garden St., Cambridge MA 02138\\
$^\dagger$
Harvard College Observatory, 60 Garden St., Cambridge MA 02138\\
${}^\ddagger$DAMTP, Silver St,Cambridge, CB3 9EW, U.K.  }
\date\today 
\maketitle

\begin{abstract}
We compute the polarization power spectra for
global strings, monopoles, textures and nontopological 
textures, and compare them to inflationary models.
We find that topological defect models predict a 
significant ($\sim 1\mu K$) contribution to magnetic type polarization 
on degree angular scales, which is produced by the large
vector component of the defect source. We also investigate
the effect of decoherence on polarization. 
It leads to a smoothing of acoustic 
oscillations both in temperature and polarization power spectra
and strongly suppresses
the cross-correlation between temperature and polarization 
relative to inflationary models.
Presence or absence of magnetic polarization or cross-correlation 
would be a strong discriminator between the two theories
of structure formation and will be testable with the next 
generation of CMB satellites.
\end{abstract}
\vskip .2in
]

\section{Introduction}
Fluctuations in the cosmic microwave background (CMB) have the promise
to become the most
powerful testing ground of cosmological models today. Their main 
advantage is that they test the universe in its early stages of
evolution, when it was much simpler than it is today. The physics that
determines the fluctuations is well understood and the 
small amplitude of perturbations allows one to use linear 
perturbation theory to perform the calculations to almost 
arbitrary accuracy. 
Moreover, theoretical predictions are 
very sensitive to various cosmological parameters, holding the
promise of their determination to a high precision.
It has long been recognized \cite{pol} 
that polarization in the microwave 
background shares these same advantages, but is sensitive 
to somewhat different physical processes and as such would 
provide a valuable complementary information to the temperature
measurements. 
 Although the amplitude of
polarization is typically less than 10\% of temperature and has not 
yet been observed so far, the sensitivity of future experiments 
such as MAP and Planck satellites
should allow one to map polarization
with a high accuracy	 
over a large fraction of the sky. 
In particular the temperature-polarisation 
cross correlation offers particular promise as a clean
observational signature 
well within reach of the next generation of CMB satellite measurements
\cite{cct1},\cite{s}. In addition, it 
was recently shown that polarization Stokes parameters
$Q$ and $U$ can be decomposed into
electric (E) and magnetic (B) components \cite{zs,kks}, 
which have opposite parities
allowing one to make a model independent identification
of non-scalar (i.e. vector or tensor) perturbations.

Most previous work on polarization has concerned the predictions of
inflationary models, where power spectra are easy 
to compute with a high accuracy, allowing one to propose several
high precision tests of cosmological models. In contrast, 
the competing theories
of cosmic structure formation, 
based on symmetry breaking and phase ordering have been
plagued by calculational difficulties preventing firm 
conclusions being drawn.
These theories involve a stiff causal source
comprising the ordering fields and/or defects, which continually
perturb the universe on ever larger scales. 
Both the nonlinear evolution of the source and a 
full linear response theory for the linearised Einstein/fluid/Boltzmann
equations are required to compute power spectra in such models. 
Recently, the first accurate calculations of power spectra in global defect
models covering  
all observational scales of interest
have been presented \cite{pst}, employing a new
two-stage calculational method. First, an
accurate numerical code for field evolution is used to measure
the unequal time
correlator of the defect source stress energy tensor $\Theta_{\mu
\nu}$. 
This quantity uses all the information present in the simulations, 
incorporates the powerful property of scaling evolution, 
and preserves enough information needed to compute 
all power spectra of interest. In the second part of the calculation, 
the unequal time correlator is decomposed into a 
sum of coherent sources, each of which can be fed into
the linearised Einstein/fluid/Boltzmann equations, 
which are evolved using a modified CMBFAST code \cite{sz}
from the early epoch until 
today. Contributions from individual coherent sources are then
added together incoherently to obtain the total power 
spectrum of interest. Several independent tests all give consistent 
results to within about 10\%.

The results of this calculation indicate that in symmetry breaking theories
a significant component of CMB anisotropies on large scales is
contributed by vector and, to a lesser extent, tensor modes, in 
addition to the usual scalar modes. The shallow peak in the power spectrum,
typically around $l \sim 100$,
is determined by the vector modes that are only weakly dependent on 
cosmological parameters. 
In addition, the defect stress-energy tensor, whose evolution is nonlinear, 
is continuously sourcing the 
metric perturbations, which in turn are sourcing the fluids. This  
leads to a decoherence: even though fluids are oscillating in 
different regions of the universe, these oscillations may be out of 
phase with each other and do not show up when averaged over the whole sky. 
Calculations
indicate that 
decoherence leads to
a partial or complete destruction of  
acoustic oscillations.  As a consequence of these two features, 
varying cosmological parameters has a smaller effect on the 
temperature spectrum than in inflationary models.

The significant amount of power seen in degree 
scale experiments compared to COBE 
already poses problems for this class of models 
\cite{pst},
although given the difficulty of these measurements 
one would require further confirmation with
the future observations before completely ruling them out.
Moreover, it is 
possible that a modified version of symmetry breaking
models that satisfies the current
observational constraints will be found and so it is
important to investigate other properties of this class of
models. In this letter we concentrate on polarization of CMB
and show that it has several characteristic properties that enable one to
distinguish the defect models from inflationary models. 

\section{Electric and Magnetic Polarization}
Polarization in the microwave background is created by
Thomson scattering of photons on electrons.
However, if the photon distribution 
function has zero quadrupole moment
in the electron rest frame then no polarization 
can be generated. Before electron-proton recombination the photon mean 
free path is very short and the system forms a perfect fluid,
whose phase space density has 
only monopole and dipole moments nonzero.
After recombination the photons start to free stream, which 
generates the quadrupole and higher moments of the distribution function.
At the same time the probability for Thomson scattering 
rapidly decreases, so polarization can only be generated during 
recombination. Its amplitude will in general be smaller than 
the amplitude of temperature fluctuations. These processes are 
generic and one expects some amount of polarization to be present 
irrespective of the specific cosmological model. However, 
symmetry breaking models differ from inflationary models in 
several aspects, of which the two most important are the relative 
contributions from scalar, vector and tensor modes and decoherence.
We will show below that both lead to a very distinctive signature 
in polarization.

Temperature and polarization spectra for various symmetry breaking models 
are shown in figure \ref{fig:fig1}. Both electric (E) and magnetic 
(B) components of polarization are shown. 
We also plot for comparison
the corresponding spectra in a typical inflationary model, which we
have taken to be the standard CDM model ($h=0.5$, $\Omega=1$, 
$\Omega_b=0.05$) with equal amount of scalars and
tensors ($T/S=1$) - the latter model has about as large 
a B component as is possible in inflationary models.
In all the models we assumed no reionization. 
The most interesting feature of the symmetry breaking models 
is the large 
magnetic mode (B) polarization, 
with a typical
amplitude of 1 $\mu K$ on degree scales. For multipoles
below $l \sim 100$ the contributions from E and B are roughly equal.
This differs strongly from the inflationary model predictions, where B is 
much smaller than E on these scales even for the extreme case of 
$T/S \sim 1$.
The reason for this difference 
is a combination of different relative
contributions from scalar, vector and tensor modes in the two classes
of models and their corresponding contributions to electric and 
magnetic types of polarization. 
Relative contributions from each type of
perturbations to T, E and B are shown explicitly in 
figure \ref{fig:fig2} for the global string model. 
Inflationary models only generate scalar and tensor modes, while 
symmetry breaking models also have a significant contribution 
from vector modes.
Scalar modes only generate E,
vector modes predominantly generate B, 
while for tensor modes E and B are comparable with B being somewhat 
smaller \cite{zs,hw}.  
Together this implies that B can be significantly
larger in symmetry breaking models than in inflationary models.

\begin{figure}
\centerline{\psfig{file=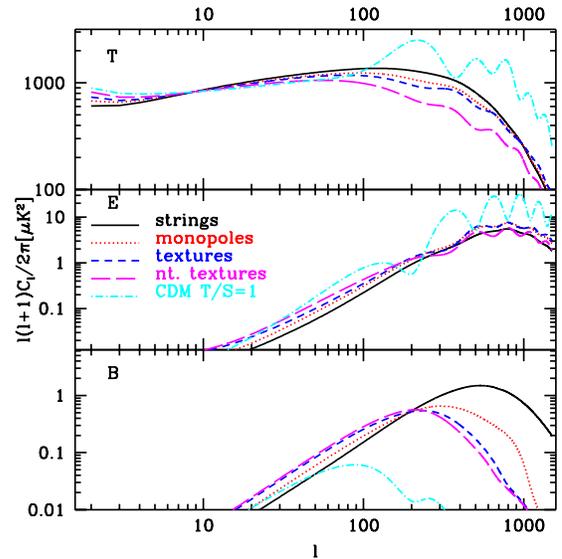,width=3.in}}
\caption{Power spectra of temperature (T), electric type polarization (E)
and magnetic type polarization (B) for global strings, monopoles, textures
and nontopological textures. For comparison we also show the corresponding
spectra in a standard CDM model with $T/S=1$ (which 
maximises the B component present). Defect models all predict 
a much larger component of B polarization on small angular scales.}
\labfig{fig1}
\end{figure}

\begin{figure}
\centerline{\psfig{file=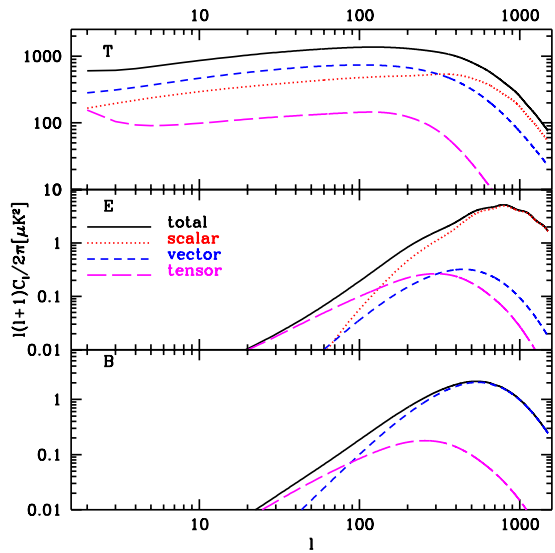,width=3.in}}
\caption{The breakdown of the contributions to the total power by the
scalar, vector and tensor components for a global string model. 
Scalars and vectors dominate E and B polarization, respectively. Other
defect models give qualitatively similar results.}
\labfig{fig2}
\end{figure}

The amplitudes of E and B polarization as a function of scale can  
be understood qualitatively from the evolution of scalar, vector 
and tensor modes.
The latter two decay away 
on sub-horizon scales, so on small angular scales only scalar 
modes are important, hence only E will contribute there in 
inflationary models. 
In such models B is typically much smaller than E on degree
scales, with the total amplitude less than 0.3 $\mu K$ in the absence 
of reionization \cite{sz97}, while the amplitude of E can be 
several $\mu K$ on sub-degree angular scales. In symmetry breaking models
most of E component is still generated by scalar modes. However, 
now B will not be negligible compared to E, because it is  
dominated by vectors, which are an important component in the defect 
source, but do not contribute significantly to E. Moreover, defects
are sourcing fluid perturbations even after horizon crossing, 
so vectors and tensors are important also on sub-degree scales.

\section{Decoherence}
Another interesting question is how coherent are polarization spectra 
in symmetry breaking models relative to their inflationary counterparts 
or to the temperature spectra.
In general one expects some degree of decoherence in any 
symmetry breaking model \cite{pst,albrecht} and this leads to 
a smearing of characteristic acoustic peaks in the spectrum. 
Appearance
of such peaks is the key requirement for the accurate determination 
of cosmological parameters, because their amplitude and position 
depends sensitively on parameters such as 
baryon and matter density, Hubble constant, curvature etc. 
Temperature spectra
in symmetry breaking model show little or no evidence of acoustic
oscillations (figure \ref{fig:fig1} and \cite{pst}). 
Both velocity and density contribute to temperature and are
out of phase in the
tightly coupled regime \cite{tc}, which leads to partial cancellation
of the peaks even before decoherence. On the other hand,
polarization receives contribution only from velocity of photon-baryon
plasma during recombination \cite{tc}, so for a coherent source
the peaks in polarization 
will be narrower than in temperature.
Note that 
acoustic oscillations are only expected for scalar modes and hence for 
E polarization. 
The spectra in figure \ref{fig:fig1} confirm this 
expectation and acoustic peaks  
are indeed somewhat more visible in E polarization than in temperature 
spectra.  
However, decoherence still plays an important role leading 
to a suppression of peaks, being progressively more important 
for lower N (where N is the dimension of the field).
In the case of strings (figure \ref{fig:fig2}) acoustic oscillations are 
completely washed out
in temperature and almost nearly so in polarization power spectra.

\begin{figure}
\centerline{\psfig{file=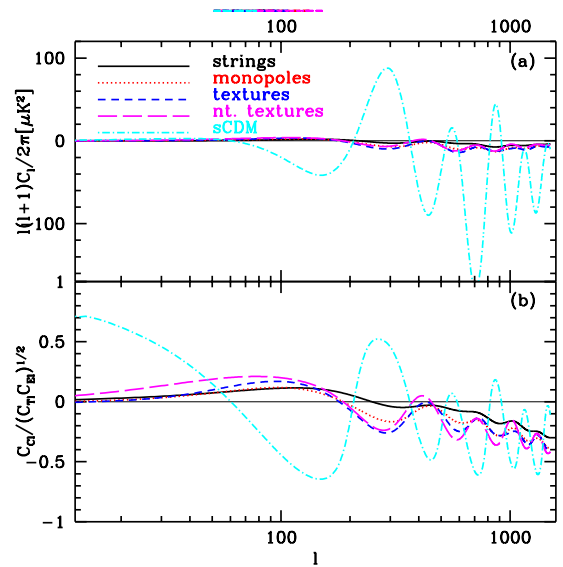,width=3.in}}
\caption{The cross-correlation power spectrum (a) and the 
correlation coefficient (b) for the same defect models as in figure 1, 
as well as for the standard CDM model.
Defect models predict significantly less power in cross-correlation 
than inflationary models.}
\labfig{fig3}
\end{figure}

Decoherence has an even more dramatic effect on the cross-correlation
between temperature and E polarization. Here the spectrum can be either
positive or negative, so decoherence may
actually destroy the cross-correlation \cite{bh}. 
In addition, causality requires the 
correlation to vanish on scales larger than the horizon in defect models 
\cite{spz}.
Figure \ref{fig:fig3}a shows the cross-correlation power spectra
for the defect models 
studied here, together with the CDM model for comparison. As expected,
cross-correlation is strongly suppressed in the defect models
relative to the inflationary model, having 5-10 times less
power. Part of this suppression is simply due to a 
smaller amplitude of both temperature and 
polarization in symmetry breaking models.
To correct for that we show in figure 
\ref{fig:fig3}b the correlation coefficient ${\rm Corr}_l=C_{Cl}/
(C_{Tl}C_{El})^{1/2}$. 
On intermediate
scales ($l<400$) cross-correlation in coherent models 
oscillates roughly around zero. Here decoherence can destroy 
any correlations; in practice decoherence is not perfect and 
some correlations remain, but the correlation coefficient is very 
small compared to inflationary models. On smaller scales the
cross-correlation becomes predominantly negative with the more
incoherent
model (strings) 
tracing the broad-band average of the more coherent ones.
The cross correlation in the defect theories is 
phase shifted by $ \sim 180^o $ relative to that in the inflationary 
model,
a result of the well known $ \sim 90^o$ phase shift in isocurvature
perturbation modes relative to adiabatic ones.
Note also the sign of the 
cross-correlation at small $l$: positive for scalar and vector perturbations 
and negative for tensors 
\cite{cct,hw}. 

\section{Discussion}
To address the question of whether the polarization 
signal discussed in this letter is detectable with the future 
CMB missions we will use a simple signal to noise estimator
\cite{knox}
\begin{equation}
{\rm S/N}=\sum_l { (2l+1)f_{\rm sky}C_l^2 \over 2[C_l+w^{-1}e^{l(l+1)\sigma_b^2}]},
\end{equation}
where $f_{\rm sky}$ is the sky coverage, $\sigma_b$ is the gaussian 
beam size of the instrument in question 
and the instrument noise is characterized with 
$w^{-1}=4 \pi \sigma^2/N$. Here $N$ is the number of pixels
and $\sigma$ is the receiver noise in each pixel. 
Typical values for MAP are $(0.1 \mu K)^2$ and $(0.15 \mu K)^2$ 
for temperature and polarization, respectively, while for Planck 
they are factor of $\sim 100$ smaller. The expression above 
includes both noise and cosmic variance and is valid for $T$, $E$ 
and $B$ power spectra, while it is somewhat more complicated for 
the cross-correlation \cite{zs,kks}. Using the typical numbers for 
MAP we 
find that it will reach S/N of order unity on both E and B 
polarization in symmetry breaking models. For an unambiguous 
detection of polarization one would therefore require either several 
years of MAP observations or the Planck mission. In the latter case
S/N will be above 20 in all the models, making possible a clear 
detection of magnetic polarization. In contrast, in inflationary 
models $S/N \sim 3$ in B polarization even if $T/S=1$. 

The prospects of detecting cross-correlation between temperature and 
polarization are even better.
The amplitude of 
cross-correlation in inflationary models is quite large and
should be detected by the MAP 
satellite over a range of scales up to $l \sim 500$ 
\cite{s}. Even for symmetry breaking models the
detection of cross-correlation 
should be marginally achievable at the MAP sensitivity levels, 
with typical ${\rm S/N} \sim 5$. This would therefore allow 
a detailed comparison between the model predictions in 
the two competing theories.
As discussed above, MAP
detection of a large signal in cross-correlation
would not favour the 
symmetry breaking models such as the ones studied here.

In conclusion, symmetry breaking models predict a large 
magnetic type polarization on small angular scales
and a small   
cross-correlation between 
temperature and polarization
relative to inflationary models. These predictions are quite 
robust:
detection of magnetic polarization on sub-degree scales
would demonstrate a presence of non-scalar 
perturbations, which should be negligible in inflationary models, 
while absence of significant
cross-correlation on degree scales would indicate  
causality and decoherence. Both predictions should be 
accessible to experimental verification in the near future.

We thank R. Battye for useful discussions and
M. Zaldarriaga for help in development of CMBFAST code.
Computing was supported by the Pittsburgh Supercomputer
Center and the UK CCC cosmology consortium.

\end{document}